\documentclass[final,5p,times,twocolumn]{elsarticle}
\usepackage{lineno,hyperref}
\usepackage{multirow}
\usepackage{multicol}
\usepackage{tikz}
 
\modulolinenumbers[5]

\journal{Journal of \LaTeX\ Templates}

\begin{document}

\begin{frontmatter}


\title{Social Media and Artificial Intelligence for Sustainable Cities and Societies: A Water Quality Analysis Use-case}

\author[1]{Muhammad Asif Auyb}
\author[2]{Muhammad Tayyab Zamir}
\author[1]{Imran Khan}
\author[1]{Hannia Naseem}
\author[1,3]{Nasir Ahmad}
\author[4]{Kashif Ahmad}
\address[1]{Department of Computer Systems Engineering, University of Engineering and Technology, Peshawar, Pakistan.}
\address[2]{Abbasyn University, Islamabad.}
\address[3]{University of Western Australia, Australia.}
\address[4]{Department of Computer Science, Munster Technological University, Cork, Ireland.}

\ead{kashif.ahmad@mtu.ie}


\begin{abstract}
This paper focuses on a very important societal challenge of water quality analysis. Being one of the key factors in the economic and social development of society, the provision of water and ensuring its quality has always remained one of the top priorities of public authorities. To ensure the quality of water, different methods for monitoring and assessing the water networks, such as offline and online surveys, are used. However, these surveys have several limitations, such as the limited number of participants and low frequency due to the labor involved in conducting such surveys.  In this paper, we propose a Natural Language Processing (NLP) framework to automatically collect and analyze water-related posts from social media for data-driven decisions. The proposed framework is composed of two components, namely (i) text classification, and (ii) topic modeling. For text classification, we propose a merit-fusion-based framework incorporating several Large Language Models (LLMs) where different weight selection and optimization methods are employed to assign weights to the LLMs. In topic modeling, we employed the BERTopic library to discover the hidden topic patterns in the water-related tweets. We also analyzed relevant tweets originating from different regions and countries to explore global, regional, and country-specific issues and water-related concerns. We also collected and manually annotated a large-scale dataset, which is expected to facilitate future research on the topic.

\end{abstract}

\begin{keyword}
Smart Cities\sep Sustainable Cities \sep  AI\sep NLP \sep Water Quality \sep Topic Modelling  \sep Named Entity Recognition \sep Text Classification\sep LLMs\sep 
\end{keyword}

\end{frontmatter}


\section{Introduction}
\label{sec:introduction}

Real-time monitoring and observation of resources and infrastructure is a primary task towards a resilient infrastructure and sustainable cities \cite{marinakis2015advanced}. This allows for taking appropriate recovery actions for the mitigation of risks and damages. Crowd-sourcing is one of the effective ways used for real-time monitoring and feedback on infrastructure \cite{helmrich2021opportunities}. One of the key methods, widely explored in the literature, for crowdsourcing is conducting surveys to obtain citizens’ feedback on different services, such as water quality, air quality, roads, infrastructure, and other societal challenges. These surveys can help in obtaining more detailed, contextual, and localized information \cite{nayak2019strengths}. These surveys are either conducted by asking citizens to fill in an online form or a questionnaire.  More recently, mobile applications have also been developed for conducting such surveys, where the participants were asked to install and give feedback. However, these online and in-person surveys have several limitations \cite{andrade2020limitations}. One of the key limitations of such surveys is the limited scope, which means they can cover a limited number of people as people are often reluctant to install such applications as we noticed during the COVID-19 pandemic \cite{ahmad2022global}. Moreover, it takes a lot of time to complete a survey and also needs to involve several human and other resources, thus, the frequency of such surveys is generally very low as it is costly and unfeasible to frequently conduct such surveys. 

These limitations of the current crowd-sourcing methods could be overcome by extracting information from social media outlets, such as Twitter and Facebook. Social media outlets have already been proven to be an effective source of communication and information spreading \cite{ahmad2019social,westerman2014social}. Their capabilities to engage large volumes of audiences worldwide make them a preferred platform for discussing and conveying concerns over different domestic and global challenges. The literature already reports their effectiveness in a diversified set of societal, environmental, and technological topics \cite{ahmad2022social,lopez2019challenges}. 

In this work, we explore the potential of social media as a crowd-sourcing source/medium of instant feedback on water quality. To this aim, an automatic solution is proposed that is able to collect and analyze citizens’ feedback on water quality. The proposed system will not only engage a large number of participants, which is a key factor in meaningful feedback, but it will also be a continuous process and will keep collecting people’s feedback continuously. One of the key advantages of the system is collecting and analyzing feedback without asking the citizens to fill in any online form/survey rather it will keep filtering and analyzing relevant social media posts in a privacy-preserving manner. 

The proposed system is composed of (i) a crawler, which is responsible for collecting social media posts (Tweets), (ii) a classification framework employing several NLP algorithms in a merit-based fusion to differentiate between water-related and irrelevant tweets, and (iii) topical analysis to automatically analyze and extract key water-related issues discussed in the tweets. For the training and evaluation of the text classification framework, we also collected and annotated a large-scale benchmark dataset, which will be made publicly available for further research in the domain.  

The key contributions of the work can be summarized as follows:

\begin{itemize}
    \item We propose an automatic tool to collect, analyze, and extract meaningful information from social media posts as a source of instant feedback on water quality, as a first step towards a sustainable water network.
    \item We propose a merit-based fusion framework by combining several transformers-based NLP algorithms to differentiate between water-related and irrelevant tweets.
    \item We also collected and annotated a large-scale benchmark dataset containing around 8,000 tweets.
    \item We also perform topic modeling on the relevant tweets to automatically extract key water-related issues discussed in the relevant tweets.
    \item We also analyze the origin of the water-related tweets and provide region and country-wise distribution of the water-related tweets collected by our system. This analysis shows the growing concern over this important societal challenge. 
\end{itemize}

The rest of the paper is organized as follows. Section \ref{sec:related_work} provides an overview of the related work. Section \ref{sec:methodology} discusses the proposed methodology. Section \ref{sec:results} covers the experimental setup, conducted experiments, and experimental results. Finally, Section \ref{sec:conclusion} concludes the paper.


\section{Related Work}
\label{sec:related_work}
The literature already reports several interesting crowdsourcing-based solutions, which are mostly based on offline or online surveys for infrastructure monitoring and feedback on different public services \cite{alavi2019overview}. The majority of the recent solutions rely on smartphones and other handheld devices by developing smart applications allowing users to give feedback on the infrastructure and services. For instance, Rapousis et al. \cite{rapousis2015qowater} proposed QoWater, a client-to-server architecture-based mobile application allowing mobile users to give feedback on water quality. Similarly, Santani et al. \cite{santani2015communisense} proposed CommuniSense, a mobile phone application for crowdsourcing to monitor road conditions in Nairobi. However, several challenges are associated with such applications for crowdsourcing \cite{sillberg2018challenges}. One of the key limitations of such surveys is the limited scope, which means they can cover a limited number of people as, generally, people are found reluctant to install and use such mobile applications.  A prime example of people’s reluctance to such mobile applications is observed during COVID-19 when people showed concerns over such applications in terms of privacy, difficulty in usage, and battery consumption \cite{ahmad2022global}. Moreover, it takes a lot of time to complete a survey and also needs to involve several human and other resources, thus, the frequency of such surveys is generally very low as it is costly and unfeasible to frequently conduct such surveys. These challenges could be overcome by extracting people's feedback on infrastructure and public services. The literature already provides some hints on the effectiveness of social media for real-time monitoring and instant feedback on different services. For instance, Want et al. \cite{wan2014improving} explore the potential of social media as a source of feedback on government services by analyzing citizens' opinions in a social media text. 

Water quality analysis is one of the key applications that recently got the attention of the community. To this aim, several interesting frameworks have been introduced. The majority of the existing works aim at sentiment analysis of social media posts to extract people's opinions on water quality. For instance, Lambert \cite{lambert2021evaluating} proposed a sentiment analysis framework for analyzing users' feedback and perception of tap water quality. Similarly, Li et al. \cite{li2021public} performed sentiment analysis on social media posts about recycled water in China. Jiang et al. \cite{jiang2016public}, on the other hand, analyzed the public's opinion on large hydro projects by performing sentiment analysis on relevant social media posts. To this aim, three different hydro projects in China are considered, and mixed opinions were noticed for the projects. 

More recently, water quality analysis from social media posts has also been introduced in MediaEval 2021 \cite{andreadis2021watermm}. The task involved the retrieval of relevant multimedia content describing water quality in an Italian region. A couple of interesting solutions, incorporating different types of available information, are proposed in response to the task. For instance, Hanif et al. \cite{HANIFwatermm} fine-tuned exiting pre-trained deep-learning models namely VGGNet and BERT for retrieving relevant visual and textual content, respectively. Overall better results are reported for textual content. Ayub et al. \cite{Asifwatermm} rather focused on textual content by employing three different NNs models including BERT, RoBERTa, and a custom LSTM both individually and jointly in a naive late fusion scheme. 

Despite the initial efforts in the domain, several interesting aspects of water quality analysis and automatic analysis of people's feedback on public services and infrastructure, in general, are unexplored. For instance, the majority of the initial efforts are based on sentiment analysis without extracting meaningful information from the content itself. The domain also lacks a large-scale benchmark dataset. To this aim, in this work, we collect and annotate a large-scale benchmark dataset on water quality analysis. We also extend the text classification framework with topic modeling to automatically extract key water-related issues discussed in social media.

\section{Methodology}
\label{sec:methodology}
\begin{figure}[]
\centering
\includegraphics[width=0.46\textwidth]{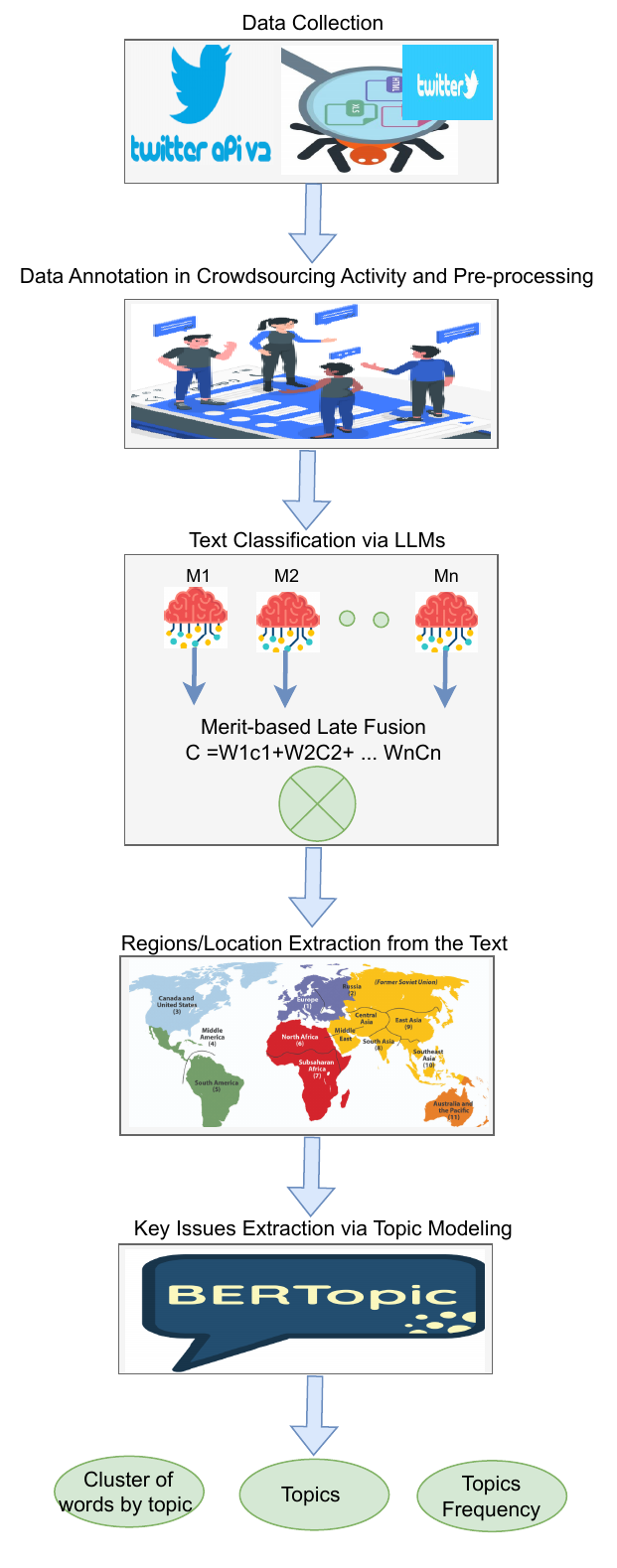}
\caption{A block diagram of the proposed methodology.}
	\label{fig:methodology}
\end{figure}

Figure \ref{fig:methodology} provides the block diagram of the proposed system. As can be seen, the proposed system is composed of five steps. In the first step, a large number of Tweets have been collected. In the next step, these tweets are annotated in a crowd-sourcing study. The annotated dataset is then used to train/finetune Large Language Models (LLMs) for the classification of tweets into relevant and non-relevant tweets. In the fourth step, several merit-based fusion techniques are used to combine the classification scores obtained with the individual models. In the final step, topic modeling techniques are used to identify topics in the relevant tweets. In the next subsections, we provide a detailed description of each step.

\subsection{Data Collection, Cleaning, and Annotation}
For data collection, we developed a crawler able to continuously collect data from different outlets of social media. As proof of concept, in the current implementation, data is collected from Twitter only.  To this aim, we used a Python package namely Tweepy\footnote{https://www.tweepy.org/} with different relevant keywords, such as \textit{waterpollution}, \textit{water}, \textit{watercrisis}, \textit{watersmell}, \textit{drinkingwater}, \textit{watercolour}, \textit{cleanwater}, \textit{waterquality}, \textit{plasticpollution}, \textit{drinkingwater}, \textit{watercrisis}, \textit{savewater}, \textit{waterislife}, \textit{cleantheocean}, \textit{plasticocean}, \textit{endplasticpollution}. The list of keywords is prepared in a data-driven manner by picking the keywords used in social media posts, blogs, newspapers etc., . We tried to include as many as possible keywords to the list to collect relevant and quality tweets. This resulted in a large collection of tweets, which were saved in a CSV file. After data collection, all the collected tweets are manually annotated by involving multiple volunteers in a crowd-sourcing activity. Before the annotation, the collected data is manually checked to remove less informative tweets. For example, we removed very short tweets without sufficient text or containing tags only. We also removed duplicate entries in the file. 

During the crowd-sourced activity, we manually analyzed a total of 8,000 tweets, which are annotated as relevant or non-relevant. To ensure, the quality of the annotated data, each sample is checked by three different annotators and is labeled based on the majority votes. The participants of the crowd-sourcing activity are postgraduate students with sufficient knowledge of the domain.  


\subsection{Text Classification}
For text classification, we employed several LLMs both individually and jointly in a merit-based fusion technique to differentiate between relevant and non-relevant tweets. In the next subsections, we provide a detailed description of the classification and fusion process.  

\subsubsection{Classification Via Individual Models}
In this work, we mainly rely on state-of-the-art transformer-based NLP models for the classification of tweets. In total, six different models are used. These models include the original BERT model, RoBERTa, ALBERT, DistilBERT, GPT, and Meta-LLAMA. The selection of these models is motivated by their proven performances in similar tasks, and we believe the evaluation of these models will provide a baseline for future work in the domain. A brief overview of these models is provided below.

\begin{itemize}
    \item \textbf{BERT}: It is one of the state-of-the-art NLP algorithms that have been widely used for a diversified list of NLP applications. Its ability to read/learn in both directions makes it a preferred choice in different text-processing applications. Several implementations of BERT are available. In this work, we used Tensorflow implementation. The model is composed of 12 layers and attention heads,
and 110 million parameters.  Our loss function is based on the Binary Cross entropy loss function while the Adaptive Moments (Adam) optimizer is used in the experiments. 
    \item \textbf{RoBERTa}: RoBERTa is another state-of-the-art transformer-based NLP model, and it uses self-attention for processing and generating contextualized representations of input text. One of the key advantages of RoBERTa over BERT is its training on a larger dataset and the use of a dynamic masking technique allowing the model to learn robust and generalizable representations of words. In this work, we fine-tuned the model on our dataset by using the Adam optimizer with a binary cross-entropy loss function.
    \item \textbf{ALBERT}: It is a modified version of BERT with fewer memory requirements. ALBERT has a reduced number of parameters mainly due to factorized embedding parameterization and cross-layer parameter sharing. In this first technique, the large vocabulary embedding matrix is decomposed into two small matrices, separating the size of the hidden layers from the size of the vocabulary embedding. The cross-layer parameter sharing, on the other hand, prevents an increase in the number of parameters with the depth of the model.
    \item \textbf{DistilBERT}: DistilBERT is another variant of the BERT model aiming at applications with less computational and memory requirements. The concept of knowledge distillation is adopted during pre-training allowing a significant reduction in parameters without a significant impact on the performance of the model. 
  \item \textbf{GPT}: Generative Pre-trained transformer (GPT) models represent a family of Neural Network (NNs)-based language prediction models built on the Transformer architecture \cite{radford2018improving}. These models are pre-trained on a huge volume of diverse text data. Currently, GPT is available in different versions. However, the first version of the model was introduced in 2018 by Open AI \cite{radford2018improving}. In this work, we used GPT version 3.5 turbo. It is composed of 175 billion parameters, which is significantly higher than the number of parameters used in its previous versions and other transformers, such as BERT. In this work, we used prompt engineering for the classification of tweets through GPT 3.5.
   \item \textbf{Meta-LLAMA}: Large Language Model Meta AI (LLMA) is also a family of pre-trained LLMs. Similar to GPT, multiple versions of LLAMA are available having 7B to 70B parameters. In this work, we used LLAMA 2, which is an improved version of the base model LLAMA. Similar to the base model, LLAMA 2 is built on the Google transformer architecture with several interesting changes and improvements. For example, the RMSNorm pre-normalization, a SwiGLU activation function, and multi-query attention instead of multi-head attention and AdamW optimizer. The key differences between LLAMA 2 and the original LLAMA include a higher context length (i.e., 4096 compared to 2048 tokens) and grouped-query attention instead of multi-query attention. Similar to GPT 3.5, we used the prompt engineering method for text classification with LLAMA.
\end{itemize}

\subsubsection{Fusion of the Models}
Our fusion methods are based on a late fusion scheme, where the scores/posterior probabilities of the individual models are accumulated for the final decision using equation \ref{eqn:fusion}. In the equation, $S_{m1}, S_{m2}, S_{m3}, ... S_{mn}$ represent the scores/posterior probabilities obtained through the $1$st, $2$nd, , $3$rd, and $n$th model, respectively while $W_{1}, W_{2}, W_{3}, ... W_{n}$ are the corresponding weights assigned to these models. 

\begin{equation}
\label{eqn:fusion}
S_{f}=W_{1}S_{m1}+W_{2}S_{m2}+W_{3}S_{m3}+....+W_{n}S_{mn}
\end{equation}

The weights are assigned to the models on the basis of their performances. To this aim, several weight optimization/selection methods, including PSO, Nelder Mead, BFGS, and Powell method, are employed. These methods seek a set of variable values (i.e., $W_{1}, W_{2}, W_{3}, ... W_{n}$ in our case) optimizing an objective function under a set of constraints. In this case, the fitness/objective function is based on accumulative classification error obtained on a validation set using equation \ref{fitness_function}. In the equation, $A_{acc}$ represents the accumulative accuracy computed on the validation set. In this work, our goal is to find a set of weights to be assigned to the models that minimize the classification error. 
\begin{equation}
e = 1-A_{acc}
	\label{fitness_function}
\end{equation}
We note that the same fitness function is used by all the weight optimization methods employed in this work. These methods use different mechanisms and have their own pros and cons. A brief overview of each method is provided below.
\begin{itemize}
    \item \textbf{PSO}: Particle Swarm Optimization (PSO), which is a heuristic approach, has been widely used in the literature for different tasks. For instance, in several works, PSO has been used for the optimization of hyper-parameters of ML algorithms, such as the number of layers, batch size, number of neurons, etc., in LSTMs and CNNs \cite{yang2020hyperparameter,kim2021optimizing}. Similarly, it has been also used for the hyper-parameter optimization of Federated Learning (FL) algorithms \cite{qolomany2020particle}. The literature also reports the effectiveness of the optimization technique in late fusion where the algorithm is used to assign optimal weights to the classifiers \cite{ahmad2018ensemble,qaraqe2020automatic}. The algorithm solves the optimization problem in three steps, iteratively; starting from a random set of candidate solutions, where each candidate solution is called a particle. At each iteration, each particle keeps track of its personal and global best solution in the swarm. The particles adjust two parameters namely (i) velocity and (ii) the position. The velocity of a particle is adjusted based on its own experience and the information shared by the other particles in the swarm. The position of particles is adjusted based on their current position, velocity, and distances between their current positions and personal and global best. The process continues until a global optimum is obtained. The key limitations of the method include a slow convergence rate, especially in high dimensional problems, and entrapment in local minima. Being one of the key optimization algorithms, PSO implementation is available in several libraries. In this work, we used the open-source library namely pyswarm\footnote{https://pyswarms.readthedocs.io/en/latest/} for the implementation of the algorithm.
    
    \item \textbf{Nelder Mead Method}: Similar to PSO, the Nelder Mead method has also been widely explored for different optimization tasks. For instance, Takenaga et al. \cite{takenaga2023practical} employed the method for computationally expensive optimization problems. Similarly, Ozaki et al. \cite{ozaki2017effective} used the algorithm for the hyper-parameter selection/optimization of a CNN model. The method has also been widely used for the fusion of classification algorithms in different visual and NLP applications \cite{zamir2023document,ahmad2022social}. The method optimizes a set of variables leading to a minimum or maximum value of an objective function in a multidimensional space. To this aim, it uses a set of $n + 1$ test points (solutions), which are arranged as a simplex. The method then estimates the behavior of the objective function at each test point for new test points, which replace the old ones in an iterative manner. In this work, we used a Python open-source library, namely, SciPy\footnote{https://scipy.org/} for the implementation of the method.
    \item \textbf{Limited-memory Broyden Fletcher Goldfarb Shanno Algorithm (BFGS)}: Similar to PSO and Neldar Mead, BFGS and its variants have been proven very effective in different tasks, such as optimization hyper-parameters of deep learning models and fusion. For instance, Saputro et al. \cite{saputro2017limited} employed the algorithm for parameter estimation on a geographically weighted ordinal logistic regression model. Maria et al. \cite{shoukat2022late} employed the method along with other optimization techniques for the fusion of inducers' scores for media interestingness prediction. BFGS, which is a local search optimization algorithm, belongs to the Quasi-Newton optimization family and aims at the optimization of the second-order derivative of the objective function.  To obtain a set of optimal values, the algorithm computes the inverse of the Hessian matrix used for multivariate functions.  To this aim, the algorithm approximates the inverse using a gradient that eliminates the need for inverse calculation at each step. One of the key limitations of the algorithm is its large memory requirement, and it becomes impractical to compute the inverse of the Hessian matrix with a larger number of input parameters. To overcome this limitation, several variations of the algorithms have been proposed. For instance, Limited BFGS/LBFGS \cite{yuan2010modified} is one of the variants of the algorithms with fewer memory requirements. In this work, though we don't have a large number of inputs, we used the LBFGS implementation of the method. 
    \item \textbf{Powell Method}: Powell method is another interesting optimization method that has been widely used for similar tasks. For instance, Maria et al. \cite{shoukat2022late} and \cite{ahmad2022social} employed the method for merit-based late fusion of classifiers for media interestingness and water quality analysis, respectively. Similar to PSO, several variations of the algorithm have been proposed in the literature. The algorithm seeks the local minima of the objective function. The objective function, which is a real-valid function with multiple inputs, doesn't need to be differentiable. The algorithm finds the minima in several steps starting with a random selection and evaluation of initial points/solutions. A list of parameters is then randomly selected. A subset of the initial points with minimum error is then selected as parents to produce children for the next step for the next generation. The children/new points are then evaluated in the fifth step and the process is repeated again from the third step until a global minima is found. 
\end{itemize}

\subsection{Regions Extraction}
In this phase, we define different regions based on the locations associated with the tweets. This allows us to analyze the water quality or water-related issues in different regions of the world as each region may have specific issues. We note that this step is added to facilitate in region-wise topic modeling, where we aim to extract keywords used in water-related tweets from different parts of the world. To this aim, the location addresses associated with each tweet are fed into Chat GPT to identify the corresponding countries by mapping the addresses to the respective countries. To ensure the quality of the mapping, the identified countries and the associated addresses are meticulously verified. To further enhance the accuracy of the data, we applied filtering techniques to specific locations. For example, in cases where the user's location included the address 'Florida, FL,' we replaced it with 'USA'. This replacement was applied wherever the specified keyword was encountered. As a result, we successfully extracted and verified 4707 accurate locations. The countries list is then provided to Chat GPT to expand the geographical scope by translating the unique countries into regions using Chat GPT.

\subsection{Topic Modeling}
The final component of the methodology is based on BERTopic \cite{grootendorst2022bertopic}, which is a state-of-the-art topic modeling technique. One of the key advantages of topic modeling is its ability to quickly discover the hidden topical patterns present in the data. These hidden patterns could result in meaningful insights leading to useful data-driven decisions.
In this work, we aim to automatically extract the hidden topical patterns in the water-related tweets to identify the key water-related issues and concerns expressed over the water quality in the tweets. 

The algorithm used in this work extracts topics from Tweets in three different steps starting from converting the tweets into embeddings, then reducing the dimensionality and clustering, and finally converting them into topics. The embeddings are obtained by a pre-trained model namely Sentence-BERT. The dimensionality reduction and clustering are carried out through Uniform Manifold Approximation and Projection(UMAP) and HDBSCAN (Hierarchical DBSCAN), respectively. Finally, topics are extracted from the clustering using a modified form of TF-IDF (Term Frequency-Inverse Document Frequency) namely c-TF-IDF.

The algorithm brings several advantages. For instance, it clusters documents based on both lexical and semantic similarities. Moreover, BERTopic provides a library with several packages allowing more accurate and better 
visualization of the clusters, topics, and probabilities. It also comes with a few limitations. For instance, its assumption that each document/tweet contains only one topic is its main limitation, though it is possible to have Tweets with multiple topics. We note that we also performed some pre-processing in addition to the data cleaning before topic modeling. For instance, we removed short and stop words, numbers, and alphanumeric characters. This allows us to remove irrelevant frequently used words.  

\section{Experiments and Results}
\label{sec:results}

\subsection{Dataset}
\label{sec:dataset}
Our final dataset, after removing less informative tweets during the manual analysis and annotation, contains a total of 7,930 tweets. Among these, 5,728 tweets are annotated as irrelevant while the remaining 2,202 tweets were classified as relevant. The dataset has been divided into three subsets namely (i) training, (ii) test, and (iii) validation set using a ratio of 70\%, 20\%, and 10\%, respectively. The validation set is used for the computation of the classification error for the fitness function of the fusion methods. Table \ref{tab:samples} provides some sample relevant and irrelevant tweets from the dataset. 

\begin{table*}[]
\centering
\caption{Sample Tweets from the dataset.}
\label{tab:samples}
\begin{tabular}{|p{6cm}|p{6cm}|}
\hline
\multicolumn{1}{|c|}{\textbf{Relevant Samples}} & \textbf{Irrelevant Samples} \\ \hline
We have been receiving water of the worst quality from past 6 months. I want to bring this situation to your notice and solve this problem ASAP. Water is a basic need. Area : Adarsh Nagar, Bahadurgarh & One of the most popular urban beaches in Gran Canaria, Las Canteras is a two-kilometer ribbon of sand caressed by warm and calm water. \\ \hline
Drinking contaminated water can transmit diseases and back in 2017 nearly 1.6 million people died from diarrheal diseases.  1/3 of those were children under the age of 5. \#climatecrisis \#water & Wondering about \#books about \#water sports (canoeing, sailing, yachting, scuba diving, etc.)? 
Check out call number range \\ \hline
The landmark research blames chemical \#pollution from plastics, farm fertilisers and pharmaceuticals in the \#water. Previously, it was thought the amount of \#plankton had halved since the 1940s, but the \#evidence gathered by the Scots suggests 90\% has now vanished. &  Hope people leave water out in their gardens or balcony in any containers for all the beautiful wildlife x \#water \#wildlife \#thirsty \#animals x \\ \hline
In face of recurring drought, cities seek security in wastewater recycling projects
 \#security \#projects \#recycling \#wastewater \#water Removing pollution from water using water shaping tech
\#sketchup \#depollution \#watershaping \#waterpollution & In a larger portion of cases, \#carpet \#damage is treated efficiently and all the defects are repaired. Professional services take care of all the \#Water \#Damage \#Restoration Sunshine Coast. \\ \hline
The privatisation of water and power has been one of the biggest rip-offs of the British public in modern times. Time to jail those profiteering through pollution of our rivers and waterways! 
\#water \#corporategreed
\#utilities  &  The theme this time is "Water from Japanese restaurants". Is it true that there are many paid shops outside Japan? The popular article has exceeded 650pv Is water free at Japanese restaurants?\\ \hline
Life without water is impossible. Save water. Save life. With every little drop, a day less to live on Earth. & Your body depends on \#water to survive. Every cell, tissue, and organ in your body needs water to work properly and for overall good health. Learn how to ensure you stay hydrated, and why it is important to do so, here in familydoctor \\ \hline
 Drinking contaminated \#water can be harmful to one's health. \#Cholera, \#diarrhea, \#dysentery, and \#typhoid are just a few of the ailments it can induce. & We've worked with TheMixUK to explain what support is available for those struggling to pay for the increasing price of \#Fuel and \#Water bills. Take a read of the article here \\ \hline
Clean Water is a necessity to daily life. Empower economically disadvantaged small communities to develop and sustain clean water supplies. & Visit Central Florida Water Ski Sweepstakes
 \\ \hline
\end{tabular}
\end{table*}

\subsection{Experimental Results}
\label{sec:results_experiments}
The objectives of this work are multi-fold. On one side, we aim to extract flood-related tweets, and on the other hand, we want to automatically extract keywords from the relevant tweets. It is very possible that each country/region may have different water-related issues than others, thus, we are also interested in keywords/topics of tweets tweeted from a specific country/region. To achieve these objectives, we perform the following experiments.
\begin{itemize}
    \item Evaluation of the performance of several state-of-the-art LLMs individually.
    \item Fusion of the classification scores obtained through the individual models in a merit-based fusion framework by employing several weight selection/optimization methods.
    \item Topic modeling on all the relevant tweets. This will allow us to highlight key global water-related issues.
    \item Topic modeling of the collection of tweets tweeted from a specific country/region. This will allow us to highlight the water-related issues specific to a particular region.
\end{itemize}
In the next subsections, we provide a detailed analysis of the results of all the experiments.

\subsubsection{Text Classification Results}
Table \ref{tab:individual_results} provides the results of our first experiments, where we evaluate the performance of several LLMs in the application. We note that for GPT and LLAMA-2 we use the prompt engineering method with a few-shot (5-shot and 10-shot) classification setting without fine-tuning the models. As can be seen, overall similar results are obtained for BERT and its different variants and XLNET. However, the lowest results are observed with Meta-LLAMA-2. One of the potential reasons for the lowest performance of the model is the few-shot learning as the model may have limited generalization to classify the samples from the seen examples.

\begin{table}[]
\centering
\caption{Experimental results of the individual LLMs.}
\label{tab:individual_results}
\begin{tabular}{|l|l|}
\hline
\multicolumn{1}{|c|}{\textbf{LLM}} & \textbf{F1-score} \\ \hline
BERT &  0.7686\\ \hline
ALBERT & 0.7636 \\ \hline
DistilBERT & 0.7491 \\ \hline
ROBERTA & 0.7541 \\ \hline
XLNET& 0.76 \\ \hline
Gpt-3.5 (5-shot)& 0.7146 \\ \hline
Gpt-3.5 (10-shot)& 0.7246 \\ \hline
Meta-LLAMA2 (5-shot)& 0.5832 \\ \hline
Meta-LLAMA2 (10-shot)& 0.5876 \\ \hline
\end{tabular}
\end{table}

Table \ref{tab:fusion_results} reports the results of our fusion experiment, where combine the classification scores of the best-performing individual models in a merit-based fusion scheme. In this experiment, we considered two experimental settings. In the first case, we combined the classification scores obtained with the top 5 performing models including BERT, RoBERTa, DistilBERT, ALBERT, and XLNET while in the second experiment, we considered the top 2 models namely BERT and ALBERT. 

Overall there is a slight improvement in the results of the fusion compared to the best-performing individual model. Generally, the fusion results in an improvement in the F1 score, however, the less improvement in this case could be the complexity of the dataset or the fewer variations in the individual models' results. As far as the comparison of the fusion methods is concerned, no significant differences have been observed. However, the performance of all the methods is higher when the top 2 best-performing models are used in the fusion compared to the top 5 models. In the case of top models, though there is no significant difference, the slightly lower performance could be due to the low-performing models that could adversely affect the performance of the fusion methods.
\begin{table}[]
\centering
\caption{Evaluation of the fusion methods.}
\label{tab:fusion_results}
\begin{tabular}{|c|cc|}
\hline
\multirow{2}{*}{\textbf{Fusion Method}} & \multicolumn{2}{c|}{\textbf{F1-score}} \\ \cline{2-3} 
 & \multicolumn{1}{c|}{Top 2 (BERT and ALBERT)} & Top 5 \\ \hline
 Simple Averaging & \multicolumn{1}{c|}{0.770} & 0.7630 \\ \hline
PSO & \multicolumn{1}{c|}{0.772} & 0.770 \\ \hline
Nelder Mead Method  & \multicolumn{1}{c|}{0.776} & 0.771 \\ \hline
Powell Method & \multicolumn{1}{c|}{0.7713} & 0.7680 \\ \hline
BFGS & \multicolumn{1}{c|}{0.77} &  0.7687\\ \hline
\end{tabular}
\end{table}

\subsubsection{Location Extraction and Topic Modeling Analysis}
\begin{figure*}[hbt!]
\centering
\includegraphics[width=0.99\textwidth]{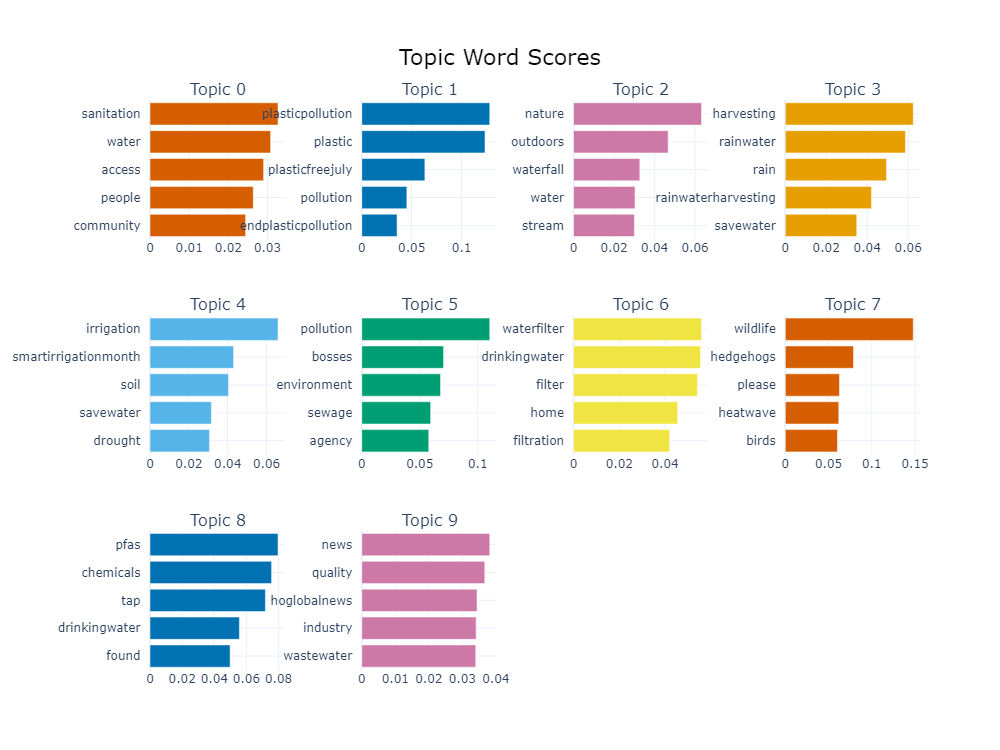}
\caption{Top 10 hidden topic patterns extracted from the complete collection of relevant tweets.}
	\label{fig:total_topics}
\end{figure*}

In the topic modeling, we conducted two different experiments. In the first case, we analyzed and tried to discover hidden topical patterns in the complete collection of water-related tweets in the test set. Figure \ref{fig:total_topics} provides the top 10 topics and the corresponding words extracted from the collection of relevant tweets through BERTopic. As can be seen, most of the topics and associated words are very relevant. The issues highlighted by the algorithm from the tweet collection include sanitation \& access to water, plastic pollution in water reservoirs, saving and utilization of rainwater, irrigation \& drought issues, environmental factors, filtering drinking water, heatwaves \& heatwave, chemicals and tap water, etc., 

\begin{figure}[hbt!]
\centering
\includegraphics[width=0.48\textwidth]{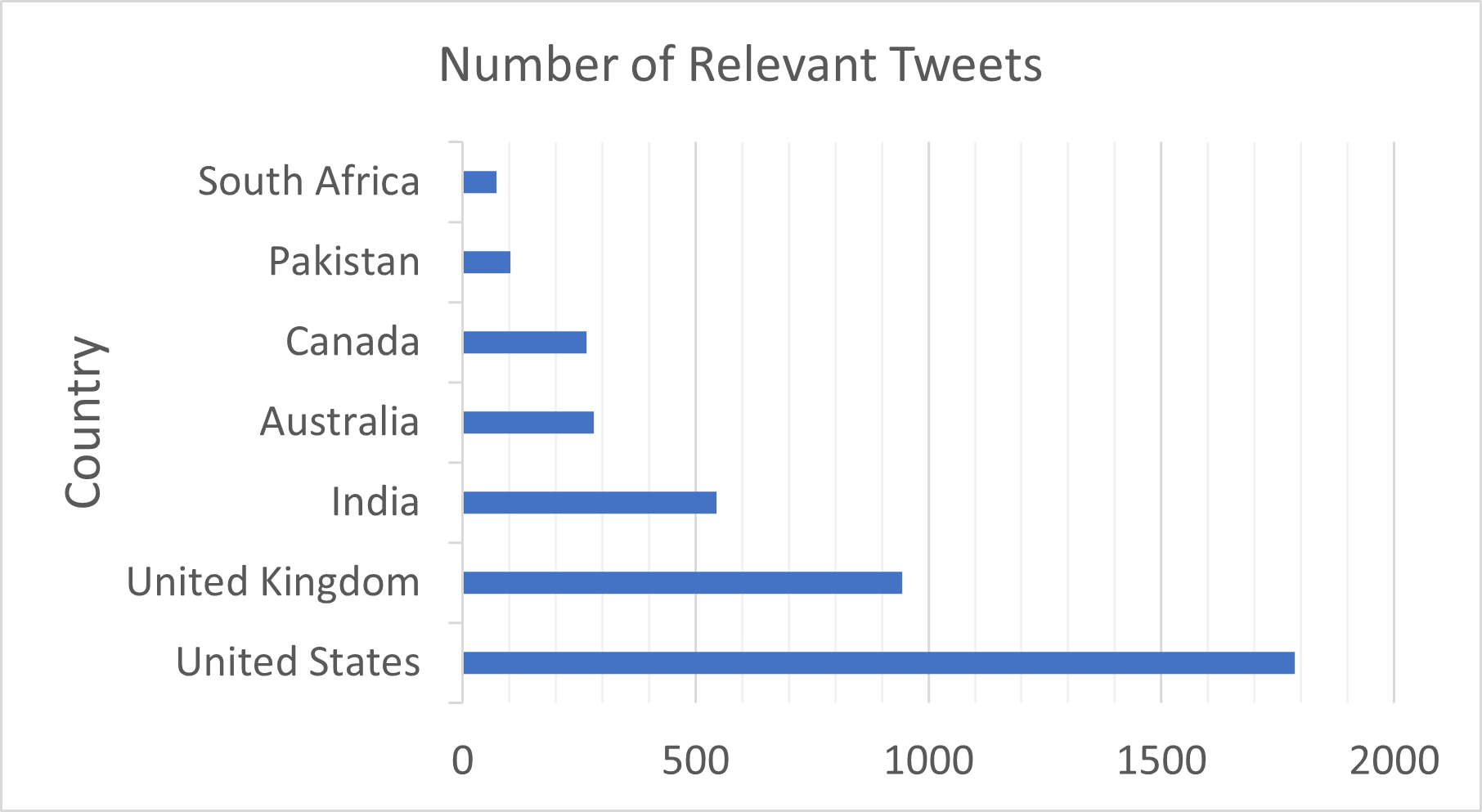}
\caption{Country-wise distribution of the origin of the water-related tweets..}
	\label{fig:country_tweets}
\end{figure}

\begin{figure*}[hbt!]
\centering
\includegraphics[width=0.99\textwidth]{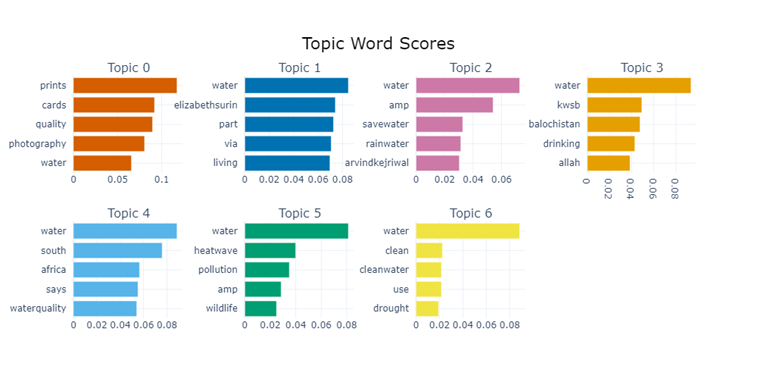}
\caption{Country-wise topic modeling of the relevant tweets. Topic 0 to topic 6 represents topics extracted from tweets originating from Australia, Canada, India, Pakistan, South Africa, the United States, and the United Kingdom, respectively.}
	\label{fig:country_topics}
\end{figure*}

In the second experiment, the collected relevant tweets were divided into regions, which allowed us to discover topics in tweets relevant to or tweeted from certain regions. This experiment helps to discover people's concerns about this important topic of water-related issues including both local and global issues. As a first step, we extracted country names from the addresses associated with relevant tweets using ChatGPT. This resulted in a long list of countries from where water-related tweets were tweeted. We observed that very few tweets were recorded from certain countries. For example, our collection of relevant tweets contains a single tweet from Slovenia, Mozambique, El Salvador, and Grenada. To ensure a sufficient number of tweets from each country, we considered only those countries from which at least 70 tweets were tweeted. We note that there is no scientific reason behind this threshold (i.e., min 70 tweets per country), we simply wanted to make sure a sufficient number of countries in our list at the same time to ensure a sufficient number of tweets for our analysis from each country. Figure \ref{fig:country_tweets} provides the country-wise distribution of the relevant tweets in our dataset. A large portion of the tweets originated from the United States and the United Kingdom. This also indicates the interest of the people from these countries in this important societal challenge.
 Figure \ref{fig:country_topics} provides the list of topics extracted from the tweets originating from different countries. Topic 0 to Topic 6 show the group of topics extracted from the tweet collections for Australia, Canada, India, Pakistan, South Africa, the United States, and the United Kingdom, respectively. Some of the topics and the associated words are less relevant compared to the others. For example, most of the words associated with topic 0, which is extracted from tweets originating from Australia, are not very relevant to water-related issues. However, on the other side, Topics 2 to Topic 6 are very relevant and helpful in highlighting the issues. For instance, Topic 2 stresses the careful usage of water in general and rainwater in particular. Topic 3 and Topic are about drinking water in one of the provinces of Pakistan and South Africa, respectively. Similarly, Topic 5 is based on heatwaves, wildlife, and water pollution. Finally, Topic 6 also includes relevant keywords, such as clean water and droughts. 

\begin{figure}[hbt!]
\centering
\includegraphics[width=0.49\textwidth]{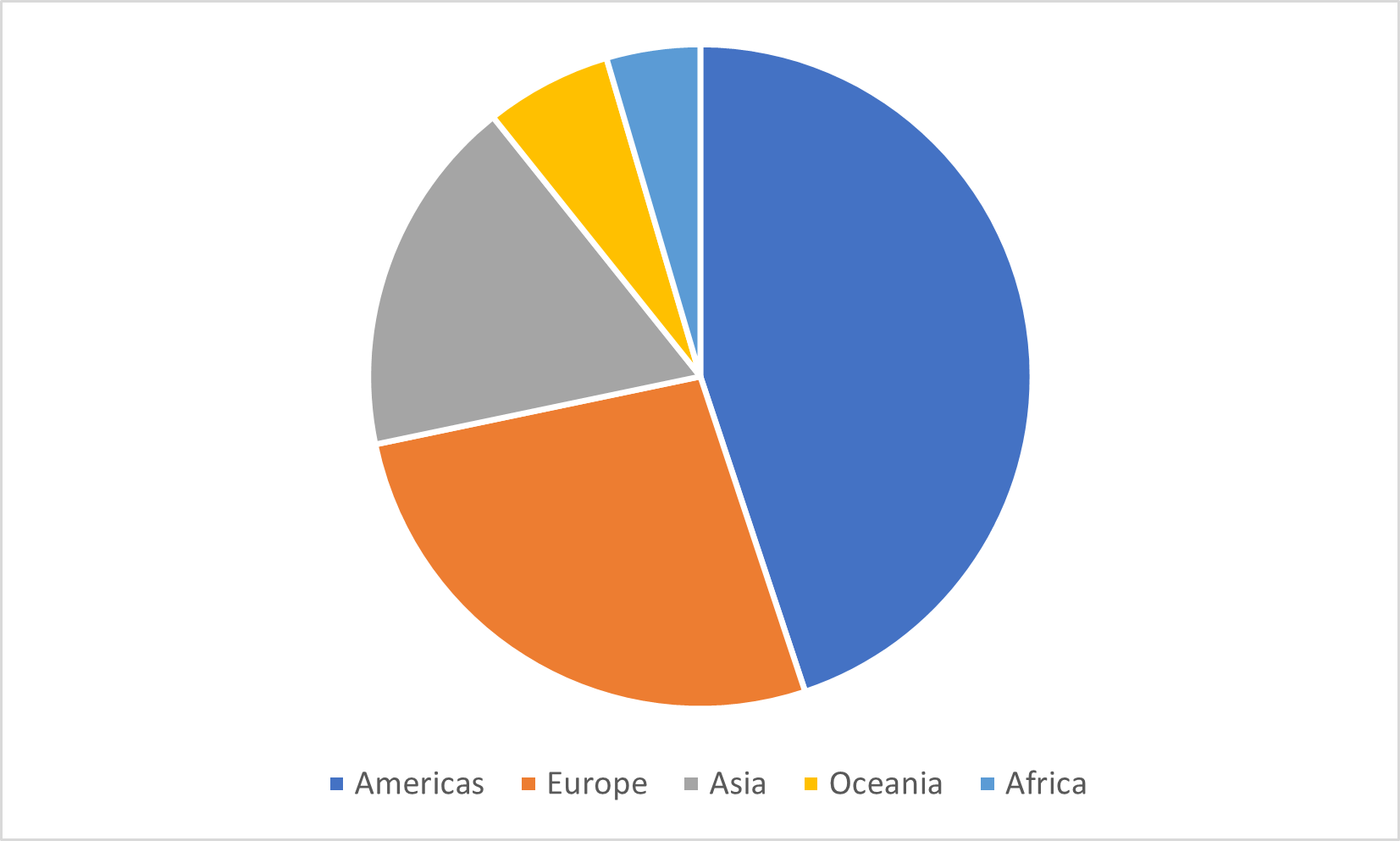}
\caption{A region-wise distribution of the origin of the water-related tweets.}
	\label{fig:region_tweets}
\end{figure}

We also performed topic modeling on different geographic regions by combining tweets from all the countries in the region. These regions are formed on the basis of the geographic locations of the countries. To this aim, the list of countries is provided to Chat GPT resulting in five regions including Asia, Africa, America, Oceania, and Europe. Similar to country-wise topic modeling, we included the regions having at least 70 tweets. Figure \ref{fig:region_tweets} provides the distribution of relevant tweets from each region. As can be seen in the figure, overall, a higher number of tweets originated from America, Europe, and Asia.   
\begin{figure*}[hbt!]
\centering
\includegraphics[width=0.99\textwidth]{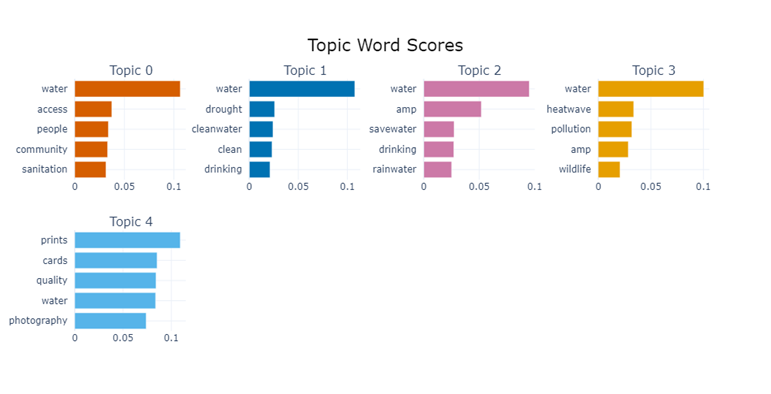}
\caption{Region-wise topic modeling of the relevant tweets. Topic 0 to topic 4 represent topics extracted from tweets originating from Africa, America, Asia, Europe, and Oceania, respectively.}
	\label{fig:region_topics}
\end{figure*}

Figure \ref{fig:region_topics} provides the summary of the topics extracted from the tweets originating from different regions. Topic 0 to topic 4 represent topics extracted from tweets originating from Africa, America, Asia, Europe, and Oceania, respectively. The majority of the topics and the associated keywords are very relevant to water quality except the topic extracted from the Oceania region. The topics are similar to what has been observed in the country-wise topic modeling, which indicates that the regions mostly have similar types of water-related issues or at least the topics/concerns are similar.

\section{Conclusion}
\label{sec:conclusion}
In this paper, we presented our solution to automatic water quality analysis and identification of hidden topic patterns in water quality-related tweets. We also analyzed the tweets originating from different countries and regions allowing global and regional water-related issues and concerns. In the text classification part, we employed several LLMs both individually and jointly in a merit-based fusion framework. No significant difference has been observed in the performance of the individual models. However, Overall, better results are obtained with the models are jointly utilized. The topic modeling techniques also efficiently extracted relevant topics from the complete collection of tweets as well as tweets originating from specific regions and countries. Moreover, we observed a large portion of the relevant tweets are from certain countries, such as the United States and the United Kingdom.

\bibliography{elsarticle-template}

\end{document}